\documentclass[letter,longauth,traditabstract]{aa}

\usepackage{amsmath}
\usepackage{txfonts}
\usepackage{graphicx}
\usepackage[colorlinks=true, linkcolor=blue, citecolor=blue, urlcolor=blue]{hyperref}
\usepackage{natbib}
\bibpunct{(}{)}{;}{a}{}{,} 
\pdfoutput=1
\begin{document}

\title{The Herschel Virgo Cluster Survey VI: The far-infrared view of
  M87\thanks{{\em{Herschel}} is an ESA space observatory with science
    instruments provided by European-led Principal Investigator
    consortia and with important participation from NASA.}}

\author{
M. Baes\inst{1} 
\and
M. Clemens\inst{2}
\and
E. M. Xilouris\inst{3}
\and
J. Fritz\inst{1}
\and
W. D. Cotton\inst{4}
\and
J. I. Davies\inst{5}
\and
G. J. Bendo\inst{6}
\and
S. Bianchi\inst{7}
\and
L. Cortese\inst{5}
\and
I. De~Looze\inst{1}
\and
M. Pohlen\inst{5}
\and
J. Verstappen\inst{1}
\and
H. B\"ohringer\inst{8}
\and
D. J. Bomans\inst{9}
\and
A. Boselli\inst{10}
\and
E. Corbelli\inst{7}
\and
A. Dariush\inst{5}
\and
S. di~Serego~Alighieri\inst{7}
\and
D. Fadda\inst{11}
\and
D. A. Garcia-Appadoo\inst{12}
\and
G. Gavazzi\inst{13}
\and
C. Giovanardi\inst{7}
\and
M. Grossi\inst{14}
\and
T. M. Hughes\inst{5}
\and
L. K. Hunt\inst{7}
\and
A. P. Jones\inst{15}
\and
S. Madden\inst{16}
\and
D. Pierini\inst{8}
\and
S. Sabatini\inst{17}
\and
M. W. L. Smith\inst{5}
\and
C. Vlahakis\inst{18} 
\and
S. Zibetti\inst{19}
}

\institute{
Sterrenkundig Observatorium, Universiteit Gent, Krijgslaan 281 S9, B-9000 Gent, Belgium 
\and
INAF-Osservatorio Astronomico di Padova, Vicolo dell'Osservatorio 5, 35122 Padova, Italy
\and
National Observatory of Athens, I. Metaxa and Vas. Pavlou, P. Penteli, GR-15236 Athens, Greece 
\and
National Radio Astronomy Observatory, 520 Edgemont Road,
Charlottesville, VA, 22903-2475, USA
\and
Department of Physics and Astronomy, Cardiff University, The Parade, Cardiff, CF24 3AA, UK
\and
Astrophysics Group, Imperial College London, Blackett Laboratory, Prince Consort Road, London SW7 2AZ, UK 
\and
INAF-Osservatorio Astrofisico di Arcetri, Largo Enrico Fermi 5, 50125 Firenze, Italy 
\and
Max-Planck-Institut f\"ur Extraterrestrische Physik, Giessenbachstrasse, Postfach 1312, D-85741, Garching, Germany
\and
Astronomical Institute, Ruhr-University Bochum, Universitaetsstr. 150, 44780 Bochum, Germany 
\and
Laboratoire d'Astrophysique de Marseille, UMR 6110 CNRS, 38 rue F. Joliot-Curie, F-13388 Marseille, France 
\and
NASA Herschel Science Center, California Institute of Technology, MS 100-22, Pasadena, CA 91125, USA 
\and
ESO, Alonso de Cordova 3107, Vitacura, Santiago, Chile 
\and
Universit\`a di Milano-Bicocca, piazza della Scienza 3, 20100, Milano, Italy 
\and
CAAUL, Observat\'orio Astron\'omico de Lisboa, Universidade de Lisboa,
Tapada da Ajuda, 1349-018, Lisboa, Portugal
\and
Institut d'Astrophysique Spatiale (IAS), Batiment 121, Universite Paris-Sud 11 and CNRS, F-91405 Orsay, France 
\and
Laboratoire AIM, CEA/DSM- CNRS - Universit\'e Paris Diderot, Irfu/Service d'Astrophysique, 91191 Gif sur Yvette, France 
\and
INAF-Istituto di Astrofisica Spaziale e Fisica Cosmica, via Fosso del Cavaliere 100, I-00133, Roma, Italy 
\and
Leiden Observatory, Leiden University, P.O. Box 9513, NL-2300 RA Leiden, The Netherlands 
\and
Max-Planck-Institut f\"ur Astronomie, Koenigstuhl 17, D-69117 Heidelberg,  Germany 
}

\date{\today}

\abstract{The origin of the far-infrared emission from the nearby
  radio galaxy M87 remains a matter of debate. Some studies find
  evidence of a far-infrared excess due to thermal dust emission,
  whereas others propose that the far-infrared emission can be
  explained by synchrotron emission without the need for an additional
  dust emission component. We present {\em{Herschel}} PACS and SPIRE
  observations of M87, taken as part of the science demonstration
  phase observations of the Herschel Virgo Cluster Survey. We compare
  these data with a synchrotron model based on mid-infrared,
  far-infrared, submm and radio data from the literature to
  investigate the origin of the far-infrared emission. Both the
  integrated SED and the {\em{Herschel}} surface brightness maps are
  adequately explained by synchrotron emission. At odds with previous
  claims, we find no evidence of a diffuse dust component in M87,
  which is not unexpected in the harsh X-ray environment of this radio
  galaxy sitting at the core of the Virgo Cluster.}


\maketitle

\section{Introduction}

At a distance of 16.7~Mpc \citep{2007ApJ...655..144M}, M87 is the
dominant galaxy of the Virgo Cluster. It is one of the nearest radio
galaxies and was the first extragalactic X-ray source to be
identified. Because of its proximity, many interesting astrophysical
phenomena can be studied in more detail in M87 than in other
comparable objects (see e.g.\ \citet{1999LNP...530.....R} for an
overview). Among its many remarkable features is the several billion
solar mass supermassive black hole at its centre
\citep{1997ApJ...489..579M, 2009ApJ...700.1690G} and the prominent jet
extending from the nucleus, visible throughout the electromagnetic
spectrum. The central regions of M87, in particular the structure of
the jet, have been studied and compared intensively at radio, optical,
and X-ray wavelengths \citep[e.g.][]{1991AJ....101.1632B,
  1996A+A...307...61M, 2001A+A...365L.181B, 2001ApJ...551..206P,
  2004ApJ...607..294S, 2005ApJ...627..140P, 2007ApJ...668L..27K,
  2008A+A...482...97S, 2010arXiv1003.5334W}.

Compared to the available information at these wavelengths, our
knowledge of M87 at far-infrared (FIR) wavelengths is rather poor. A
controversial issue is the origin of the FIR emission in M87, i.e.,
the question of whether the FIR emission is caused entirely by
synchrotron emission or whether there is an additional contribution
from dust associated with either the global interstellar medium or a
nuclear dust component. This question is partly driven by the
observation of faint dust features in deep optical images
\citep{1993ApJ...413..531S, 2006ApJS..164..334F}. Several papers on
the FIR emission of M87 arrive at different
conclusions. \citet{2007ApJ...663..808P} present ground-based Subaru
and {\em{Spitzer}} IRS spectra of the M87 nucleus and find evidence of
an excess at wavelengths longer than 25~$\mu$m, which they attribute
to thermal emission from cool dust at a characteristic temperature of
some 55~K. This claim is countered by \citet{2009ApJ...705..356B}, who
present a higher signal-to-noise IRS spectrum of the nucleus. After
careful subtraction of a stellar emission template from the
mid-infrared spectrum, these authors conclude that the nuclear
spectrum can be fully accounted for by optically thin synchrotron
emission and that there is little room for dust emission. On a larger
scale, \citet{2004A+A...416...41X} present ISOCAM imaging of M87 and
argue that the mid-infrared flux can be attributed to a single
synchrotron emission component. \citet{2007ApJ...655..781S} present
{\em{Spitzer}} IRAC and MIPS imaging of M87 and find a slight excess
in the FIR over a power-law interpolation. They attribute this excess
emission to dust emission from the host galaxy. Finally,
\citet{2008ApJ...689..775T} observed 1.3~mm continuum emission from
the nucleus and jet of M87 and found that the measured fluxes are
generally consistent with synchrotron emission, although they could
not rule out a possible nuclear contribution from thermal dust
emission.

The recently launched {\em{Herschel}} Space Observatory
\citep{Herschel} offers the possibility to study M87 at FIR
wavelengths in more detail than has been possible to date. The PACS
\citep{PACS} and SPIRE \citep{SPIRE} instruments combined can produce
images over the wavelength range between 70 and 500~$\mu$m with
unprecedented sensitivity and superior resolution. In this Letter, we
present PACS and SPIRE imaging of M87 at 100, 160, 250, 350, and
500~$\mu$m, taken as part of the science demonstration phase (SDP)
observations of the Herschel Virgo Cluster Survey
(HeViCS\footnote{More details on HeViCS can be found on
  \href{http://www.hevics.org}{http://www.hevics.org}.}). We combine
these observations with mid-infrared, FIR, submm, and radio data from
the literature to investigate the level and the origin of the FIR
emission in M87. In Sect.~{\ref{Observations.sec}}, we present the
observations and data reduction, Sect.~{\ref{Analysis.sec}} describes
the analysis of the data and in Sect.~{\ref{Conclusion.sec}} we
present our conclusions.

\section{Observations and data reduction}
\label{Observations.sec}

\begin{figure}
  \centering
  \includegraphics[width=0.485\textwidth]{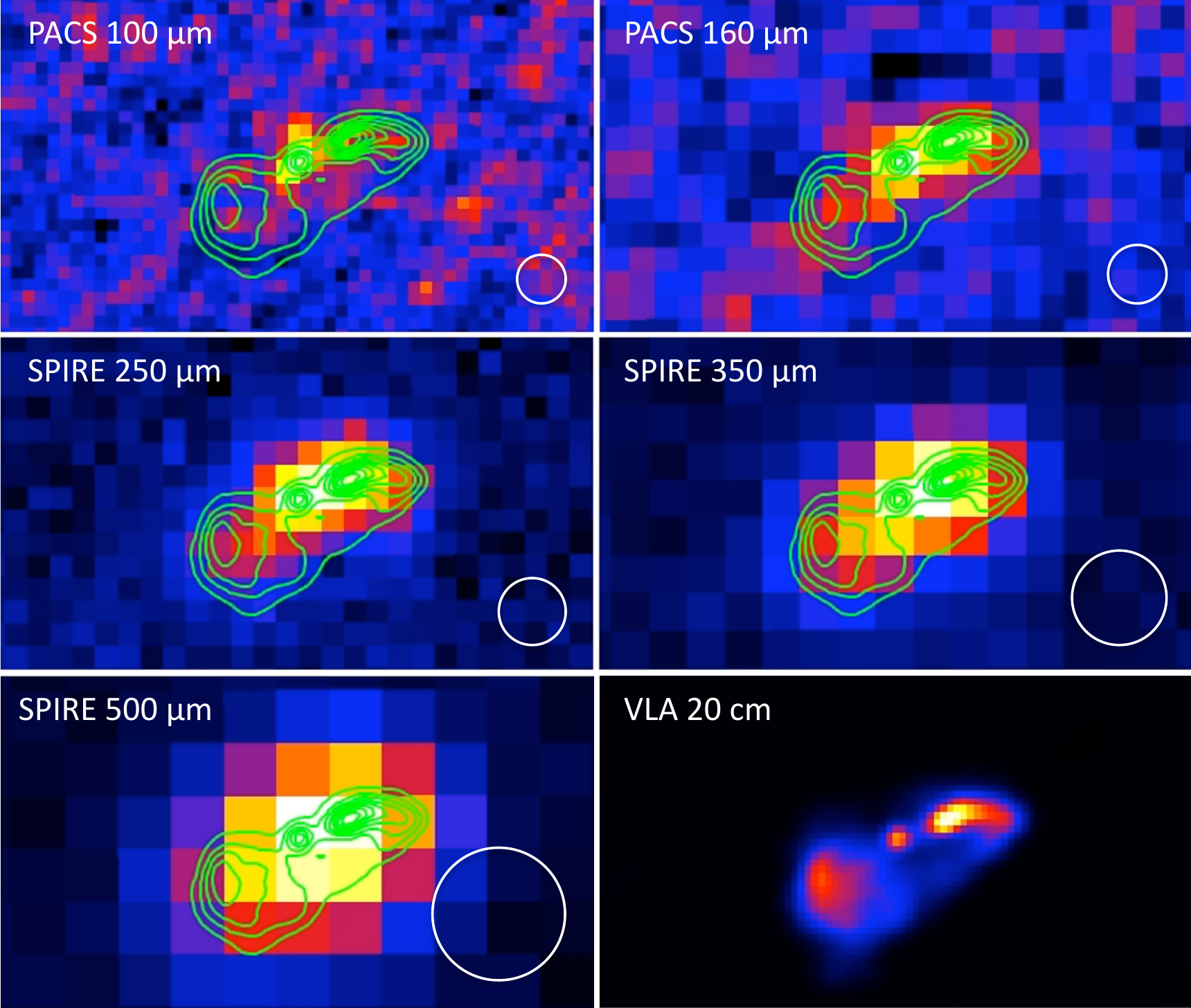}
  \caption{The {\em{Herschel}} view of the central regions of M87. The
    bottom right image is a VLA 20~cm image from the FIRST survey. The
    20~cm radio contours have been overlaid on the {\em{Herschel}}
    images. The field of view of all images is
    $160\arcsec\times90\arcsec$, beam sizes are indicated in the
    bottom right corner.}
  \label{RawData.pdf}
\end{figure}

We observed M87 on 29 November 2009 with PACS and SPIRE as part of the
HeViCS SDP observations. The HeViCS SDP field covers a
$4\times4$~deg$^2$ field at the centre of the Virgo Cluster, roughly
centred on M87. It was scanned with a 60\arcsec/s scanning speed in
nomimal and orthogonal directions. Data were gathered simultaneously
in the green and red PACS bands (100 and 160~$\mu$m) and the three
SPIRE bands (250, 350, and 500~$\mu$m). The PACS and SPIRE data were
reduced using HIPE, with reduction scripts based on the standard
reduction pipelines. For more details of the HeViCS SDP data
reduction, we refer to \citet{HeViCS-paper1}.
Figure~{\ref{RawData.pdf}} shows the {\em{Herschel}} images at the
five PACS and SPIRE bands of the central $160\arcsec\times90\arcsec$
regions of M87, which is clearly detected in all five bands.

\section{Analysis}
\label{Analysis.sec}

\begin{figure}
  \centering
  \includegraphics[width=0.48\textwidth]{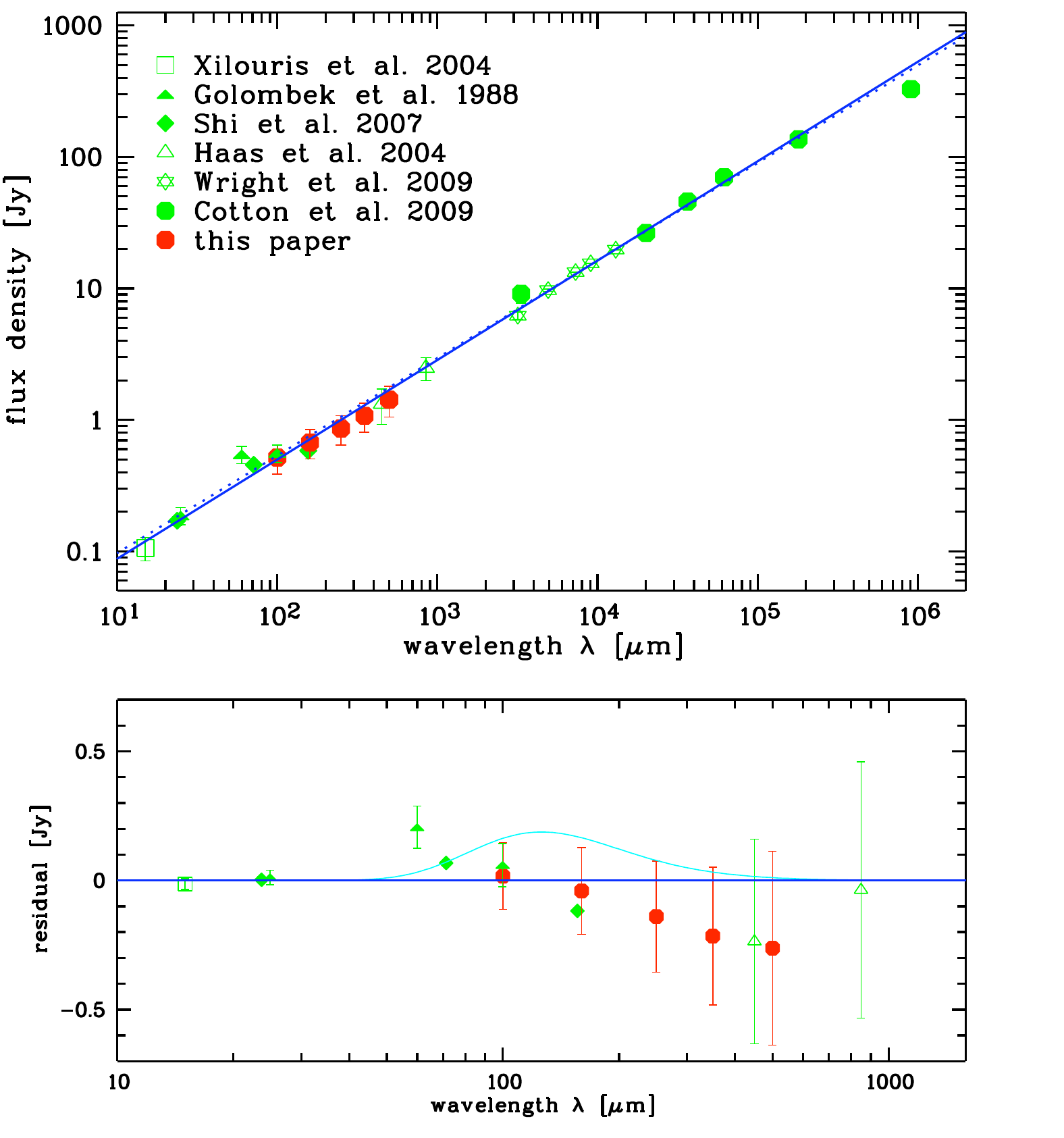}
  \caption{Top: the global SED of M87 from mid-infrared to radio
    wavelengths. When no error bars are seen, they are smaller than
    the symbol size.  The solid line in the plot is the best-fit power
    law of the ISOCAM, IRAS, MIPS, SCUBA, GBT, WMAP, and VLA data; the
    dotted line has only been fitted to the SCUBA, GBT, WMAP, and VLA
    data. Bottom: residual between data and the best-fit synchrotron
    model in the infrared-submm wavelength range. The cyan line is a
    modified black-body model with $T=23$~K and
    $M_{\text{d}}=7\times10^4~M_\odot$ (see text).}
  \label{SED.pdf}
\end{figure}
 
\begin{figure*}
  \centering
  \includegraphics[width=0.8\textwidth]{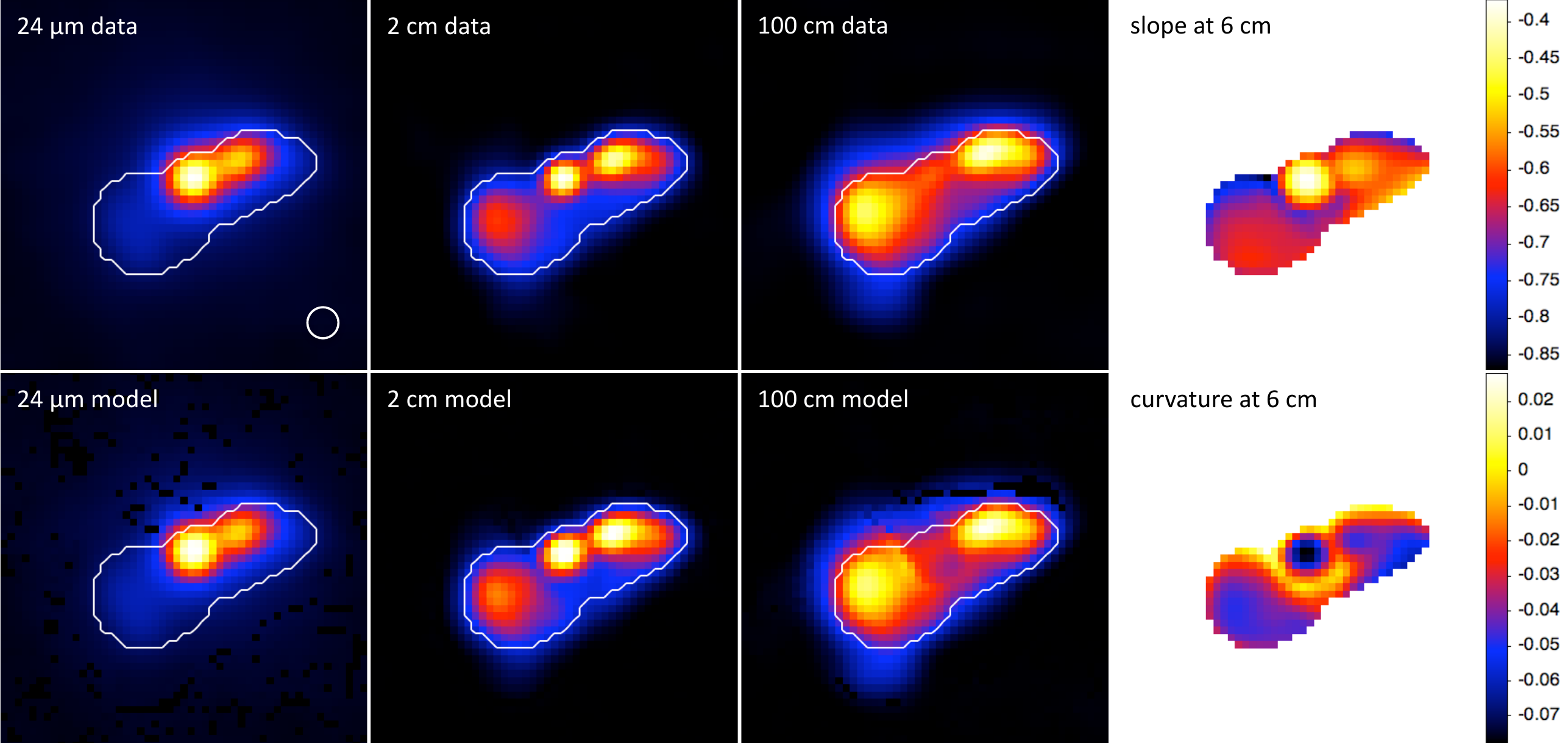}
  \caption{A comparison between the data and a synchrotron model. The
    three images on the top row show the observed {\em{Spitzer}} image
    of M87 at 24~$\mu$m and VLA images at 2~cm and 100~cm, all
    convolved to the same resolution (8.5\arcsec\ FWHM, 2\arcsec\
    pixels). In each pixel, a synchrotron emission was fit to the
    {\em{Spitzer}}, GBT, and VLA data accross the wavelength range
    between 24~$\mu$m and 100~cm. The solid lines in the images show
    the region where the data are reliable at all wavelengths:
    outside this region, not all images could be used. The bottom row
    images show the synchrotron model images corresponding to the
    images on the top row. The two rightmost images show the slope of
    the synchrotron fit and the curvature of the fit, both at 6~cm.}
  \label{SynchrotronModel.pdf}
\end{figure*}
 
The most straightforward, first-order approach to investigating the
origin of the FIR emission in M87 is to study its global spectral
energy distribution (SED). Table~{\ref{SED.tab}} lists the
{\em{Herschel}} fluxes of M87, together with ISOCAM, IRAS, MIPS, and
SCUBA data gathered from the literature \citep{2004A+A...416...41X,
  1988AJ.....95...26G, 2007ApJ...655..781S, 2004A+A...424..531H}. The
top panel in Fig.~{\ref{SED.pdf}} shows the SED in the
infrared-submm-radio region between 15~$\mu$m and 100~cm. Apart from
the infrared-submm fluxes from Table~{\ref{SED.tab}}, this plot shows
GBT MUSTANG and VLA radio continuum fluxes from
\citet{2009ApJ...701.1872C} and the most recent 5-year WMAP fluxes
from \citet{2009ApJS..180..283W}. The solid line is the best-fit power
law for the ISOCAM, IRAS, MIPS, SCUBA, GBT, WMAP, and VLA data and has
a slope $\alpha=-0.76$; the dotted line fits only the SCUBA, GBT,
WMAP, and VLA data and has a slope $\alpha=-0.74$.

The bottom panel in Fig.~{\ref{SED.pdf}} shows the residual from the
best-fit power law in the infrared-submm wavelength region; it is
clear that the integrated {\em{Herschel}} fluxes are in full agreement
with synchrotron radiation. The cyan line in this figure is a modified
black-body fitted with $T=23$~K and
$M_{\text{d}}=7\times10^4~M_\odot$. This temperature is the mean dust
equilibrium temperature in the interstellar radiation field of M87,
determined using the SKIRT radiative transfer code
\citep{2003MNRAS.343.1081B, 2005AIPC..761...27B} and based on the
photometry from \citet{2009ApJS..182..216K}. The dust mass was
adjusted to fit the upper limits of the residuals. It is clear that
the SED of M87 is incompatible with dust masses higher than
$10^5~M_\odot$.

Although indicative, the analysis of the integrated SED does not
definitely identify the origin of the FIR emission in
M87. Approximating the global SED as a single power-law synchrotron
model is indeeed an oversimplification of the complicated structure of
M87. The bottom-right panel of Fig.~{\ref{RawData.pdf}} shows a
high-resolution 20~cm image from the VLA FIRST survey
\citep{1995ApJ...450..559B,1997ApJ...475..479W}, its contours being
superimposed on the {\em{Herschel}} images. This 20~cm image
identifies three distinct regions of significant synchrotron emission:
the nucleus, the jet and associated lobes in the NW region, and the SE
lobes. It is well-known that these different components have different
spectral indices \citep[e.g.,][]{2007ApJ...655..781S,
  2009ApJ...701.1872C} -- detailed studies have shown that even within
the jet, the spectral index can vary significantly
\citep{1991AJ....101.1632B, 1996A+A...307...61M,
  2001ApJ...551..206P}. A spatially resolved analysis of the different
regions of M87 is therefore a more powerful tool for investigating the
origin of the FIR emission. Unfortunately, a full spatially resolved
analysis of M87 using all {\em{Herschel}} bands is impractical. On the
one hand, the PACS PSF is smeared significantly due to the rapid scan
speed and the PACS images have a limited signal-to-noise ratio
(hereafter S/N). The SPIRE images on the other hand have higher S/N,
but the beam size is relatively large for a full spatially resolved
analysis of the different components. The M87 nucleus, jet, and
associated NW lobes are discernible in the three SPIRE bands, but the
identification of faint emission in the SE lobes is difficult,
particularly in the 500~$\mu$m band. We therefore concentrated our
efforts on the 250~$\mu$m image, which provides the optimal compromise
between S/N and spatial resolution.

Our analysis consisted of constructing a synchrotron model for the
central regions of M87, based on the available ancillary images with
sufficient field-of-view, resolution, and surface brightness
sensitivity. We use the new 90~GHz radio continuum maps taken with the
MUSTANG bolometer array presented by \cite{2009ApJ...701.1872C}. These
data have a resolution of 8\farcs5 FWHM, hence closely match our
{\em{Herschel}} data. The same authors also present archival VLA data
at 15, 8.2, 4.9, 1.6, and 0.3~GHz of similar resolution, which were
also used in our analysis. An image at 23 GHz was also available in
the VLA archive, but it was not used in our analysis because it may be
insensitive to extended structure because of a lack of short
interferometer baselines. Finally, a {\em{Spitzer}} MIPS image at
24~$\mu$m was extracted from the archive and reduced using the MIPS
Data Analysis Tools \citep{2005PASP..117..503G}, as described in
\citet{2009AJ....137.3053Y}. As noted by \citet{2007ApJ...655..781S},
a faint extended halo is visible in the 24~$\mu$m image, which is
probably caused by stellar emission and/or circumstellar dust. Because
of the complexity of the image, we did not attempt to subtract this
faint emission from the map; since the flux density in the central
regions of M87 is strongly dominated by non-thermal emission from the
nucleus, jet, and lobes, we are confident that this decision does not
affect our results.

As a first step, we convolved the available images to the same
resolution (8\farcs5 FWHM), shifted them to the same astrometry
(2\arcsec\ pixel scale), and converted them to similar surface
brightness units (MJy\,sr$^{-1}$). The result is a data cube with
seven points in the wavelength dimension covering an impressive
wavelength range of nearly 5 orders of magnitude. Following the same
strategy as \citet{2009ApJ...701.1872C}, we fitted a second-order
polynomial synchrotron model to each pixel of this data cube.  The
fits were performed with the MPFIT robust non-linear least squares
curve fitting library in IDL \citep{2009ASPC..411..251M}.

Figure~{\ref{SynchrotronModel.pdf}} shows a comparison between the
data and the resulting synchrotron models at 24~$\mu$m, 2~cm, and
100~cm. The solid white line in these figures borders the spatial
region where reliable surface brightnesses were available at all seven
wavelengths. Outside this region, spectral fits were still made, but
typically used fewer data points, e.g., because one of the radio
images contained negative flux at the corresponding wavelengths. By
inspecting the images on the top row, it is obvious that the
synchrotron emission is not homogeneous in the different regions: the
short wavelength radiation is dominated by the central source, whereas
the jet and lobes dominate at the longer wavelengths. Comparing the
top row panels with the corresponding bottom row panels, however, one
can clearly see that, in spite of this different structure, our simple
synchrotron model can reproduce the observed images very well over the
entire wavelength range. The right-hand panels of
Fig.~{\ref{SynchrotronModel.pdf}} show the slope and the curvature of
the synchrotron model at the reference wavelength of 6~cm. The slope
varies from $-0.4$ in the nuclear region to $-0.8$, and the curvature
is modest ranging from slightly positive in the very faint regions to
about $-0.06$ in nucleus, jet, and lobes. These values are consistent
with the values found by \citet{2009ApJ...701.1872C}, who limited
their fits to the radio regime between 3.3~mm and 100~cm.

\begin{table}
\centering
\caption{Integrated fluxes for M87 in the infrared-submm wavelength region
  between 15 and 1000~$\mu$m.}
\begin{tabular}{ccll}
  \hline\hline \\
  instrument & $\lambda$ ($\mu$m) & $F_\lambda$ (Jy) & source
  \\ \\ \hline \\
  ISOCAM & 15 & $0.106 \pm 0.021$ & \citet{2004A+A...416...41X} \\
  MIPS & 24 & $0.171 \pm 0.013$ & \citet{2007ApJ...655..781S} \\
  IRAS & 25 & $0.187 \pm 0.028$ & \citet{1988AJ.....95...26G} \\
  IRAS & 60 & $0.546 \pm 0.082$ & \citet{1988AJ.....95...26G} \\
  MIPS & 70 & $0.455 \pm 0.009$ & \citet{2007ApJ...655..781S} \\
  IRAS & 100 & $0.559 \pm 0.084$ & \citet{1988AJ.....95...26G} \\
  PACS & 100 & $0.517 \pm 0.129$\footnotemark[2] & this paper \\
  MIPS & 160 & $0.582 \pm 0.010$ & \citet{2007ApJ...655..781S} \\
  PACS & 160 & $0.673 \pm 0.168$\footnotemark[2] & this paper  \\
  SPIRE & 250 & $0.860 \pm 0.215$\footnotemark[2] & this paper \\
  SPIRE & 350 & $1.070 \pm 0.267$\footnotemark[2] & this paper \\
  SCUBA & 450 & $1.320 \pm 0.396$ & \citet{2004A+A...424..531H} \\
  SPIRE & 500 & $1.430 \pm 0.375$\footnotemark[2] & this paper \\
  SCUBA & 850 & $2.480 \pm 0.496$ & \citet{2004A+A...424..531H} \\ \\
 \hline\hline
\end{tabular}
\label{SED.tab}
\end{table}

\begin{figure}
  \centering
  \includegraphics[width=0.485\textwidth]{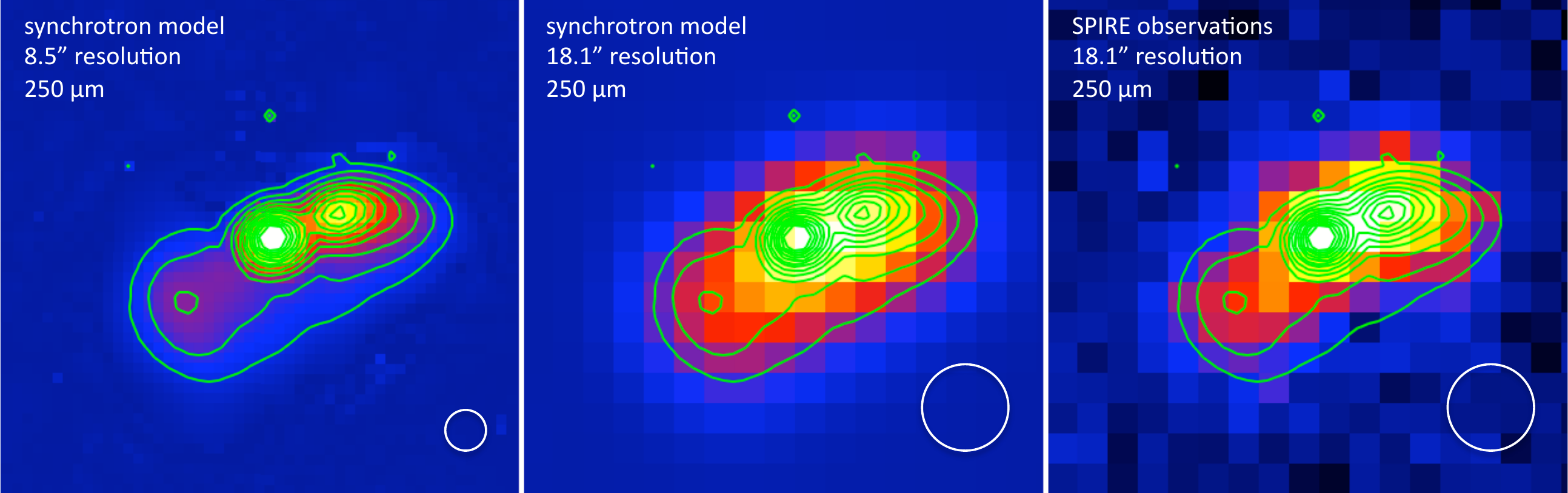}
  \caption{A comparison between the synchrotron model image and the
    observed image at 250~$\mu$m. The left panel shows the synchrotron
    image at the model resolution (8\farcs5 FWHM, 2\arcsec\ pixels),
    the central panel shows the same model convolved to the SPIRE
    250~$\mu$m beam and pixel size (18\farcs1 FWHM, 6\arcsec\
    pixels). The right panel shows the observed SPIRE 250~$\mu$m
    image. In all panels, the green lines are the contours of the
    synchrotron model at the model resolution.}
  \label{H250.pdf}
\end{figure}

For the next step in our analysis, we used our synchrotron model to
predict the emission of M87 at 250~$\mu$m. The left panel of
Fig.~{\ref{H250.pdf}} shows the synchrotron model prediction at
250~$\mu$m at the model resolution. By far the brightest peak of the
model is located at the position of the nucleus; the jet is clearly
visible as a second bright component, whereas the SE lobe is visible
as an extended, low surface brightness region with a peak surface
brightness that is almost five times fainter than the nucleus (23
MJy\,sr$^{-1}$ versus 110 MJy\,sr$^{-1}$). When we convolve this
synchrotron model image with the SPIRE 250~$\mu$m beam and rebin it to
match the observed SPIRE image astrometry, the three different
components corresponding to nucleus, jet, and SE lobes merge into a
single extended structure with one elongated peak 4\arcsec\ to the
west of the nucleus. Comparing the central and right panels of
Fig.~{\ref{H250.pdf}}, we see that the synchrotron model is capable of
explaining the observed SPIRE 250~$\mu$m image satisfactorily.

\footnotetext[2]{We assume a conservative uncertainty in
  the PACS and SPIRE flux densities of 25\%, which includes the 15\%
  uncertainty in the absolute flux calibration \citep{PACS, Swinyard},
  uncertainties introduced by the map-making techniques, and the
  possible contamination of background objects. Colour corrections
  were applied, but they turned out to be very small (1.8\% for the
  PACS\,100 band and smaller than 1\% for the remaining bands).}  

\section{Conclusion}
\label{Conclusion.sec}

 For both the integrated SED and the SPIRE 250~$\mu$m map, we have
found that synchrotron emission is an adequate explanation of the FIR
emission. We do not detect a FIR excess that cannot be explained by
the synchrotron model. In particular, we have no reason to invoke the
presence of smooth dust emission associated with the galaxy
interstellar medium, as advocated by \citet{2007ApJ...655..781S}. For
a dust temperature of 23~K, which is the expected equilibrium
temperature in the interstellar radiation field of M87, we find an
upper limit to the dust mass of $7\times10^4~M_\odot$. Our result
agrees with the analysis of the nuclear emission by
\citet{2009ApJ...705..356B}. \citet{Clemens} discuss the lifetimes of
interstellar dust grains in elliptical galaxies in the Virgo Cluster
based on {\em{Herschel}} observations and find an upper limit to the
amorphous silicate grain survival time of less than 46 million
years. Given that M87 is a luminous X-ray source, the absence of a
substantial dust component is not a surprise. A low dust content is
also in agreement with the non-detection of cool molecular gas
\citep{2008A+A...489..101S, 2008ApJ...689..775T} and the non-detection
of significant intrinsic absorption in the X-ray spectra of M87
\citep{2002A+A...382..804B}. Our conclusion is that, seen from the FIR
point of view, M87 is a passive object with a central radio source
emitting synchrotron emission, without a substantial diffuse dust
component.

\begin{acknowledgement}
  The National Radio Astronomy Observatory (NRAO) is operated by
  Associated Universities Inc, under cooperative agreement with the
  National Science Foundation.
\end{acknowledgement}


\begin{thebibliography}{}

\bibitem[Baes et al.(2003)]{2003MNRAS.343.1081B} Baes, M., et al.\
  2003, \mnras, 343, 1081

\bibitem[Baes et al.(2005)]{2005AIPC..761...27B} Baes, M., Dejonghe,
  H., \& Davies, J.~I.\ 2005, The Spectral Energy Distributions of
  Gas-Rich Galaxies: Confronting Models with Data, 761, 27

\bibitem[Becker et al.(1995)]{1995ApJ...450..559B} Becker, R.~H., White, 
R.~L., \& Helfand, D.~J.\ 1995, \apj, 450, 559 

\bibitem[Biretta et al.(1991)]{1991AJ....101.1632B} Biretta, J.~A., Stern, 
C.~P., \& Harris, D.~E.\ 1991, \aj, 101, 1632 

\bibitem[B{\"o}hringer et al.(2001)]{2001A+A...365L.181B}
  B{\"o}hringer, H., et al.\ 2001, \aap, 365, L181

\bibitem[B{\"o}hringer et al.(2002)]{2002A+A...382..804B}
  B{\"o}hringer, H., et al.\ 2002, \aap, 382, 804

\bibitem[Buson et al.(2009)]{2009ApJ...705..356B} Buson, L., et al.\ 2009, 
\apj, 705, 356 

\bibitem[Clemens et al.(2010)]{Clemens} Clemens, M., et al.\ 2010,
  A\&A, this issue

\bibitem[Cotton et al.(2009)]{2009ApJ...701.1872C} Cotton, W.~D., et al.\ 
2009, \apj, 701, 1872 

\bibitem[Davies et al.(2010)]{HeViCS-paper1} Davies, J.~I., et al.\
  2010, A\&A, this issue

\bibitem[Ferrarese et al.(2006)]{2006ApJS..164..334F} Ferrarese, L.,
  et al.\ 2006, \apjs, 164, 334

\bibitem[Gebhardt \& Thomas(2009)]{2009ApJ...700.1690G} Gebhardt, K.,
  \& Thomas, J.\ 2009, \apj, 700, 1690

\bibitem[Golombek et al.(1988)]{1988AJ.....95...26G} Golombek, D., Miley, 
G.~K., \& Neugebauer, G.\ 1988, \aj, 95, 26 

\bibitem[Gordon et al.(2005)]{2005PASP..117..503G} Gordon, K.~D., et
  al.\ 2005, \pasp, 117, 503

\bibitem[Griffin et al.(2010)]{SPIRE} Griffin, M., et al.\ 2010, A\&A,
  this issue

\bibitem[Haas et 
al.(2004)]{2004A+A...424..531H} Haas, M., et al.\ 2004, \aap, 424, 531 

\bibitem[Kormendy et al.(2009)]{2009ApJS..182..216K} Kormendy, J., Fisher, 
D.~B., Cornell, M.~E., \& Bender, R.\ 2009, \apjs, 182, 216 

\bibitem[Kovalev et al.(2007)]{2007ApJ...668L..27K} Kovalev, Y.~Y., et
  al.\ 2007, \apjl, 668, L27

\bibitem[Macchetto et al.(1997)]{1997ApJ...489..579M} Macchetto, F., 
Marconi, A., Axon, D.~J., Capetti, A., Sparks, W., 
\& Crane, P.\ 1997, \apj, 489, 579 

\bibitem[Markwardt(2009)]{2009ASPC..411..251M} Markwardt, C.~B.\ 2009,
  ASP Conf.\ Ser., 411, 251

\bibitem[Marshall et al.(2002)]{2002ApJ...564..683M} Marshall, H.~L., 
et al.\ 2002, \apj, 564, 683 

\bibitem[Mei et al.(2007)]{2007ApJ...655..144M} Mei, S., et al.\ 2007, 
\apj, 655, 144 

\bibitem[Meisenheimer et 
al.(1996)]{1996A+A...307...61M} Meisenheimer, K., Roeser, H.-J., \&
Schloetelburg, M.\ 1996, \aap, 307, 61

\bibitem[Owen et al.(2000)]{2000ApJ...543..611O} Owen, F.~N., Eilek, J.~A., 
\& Kassim, N.~E.\ 2000, \apj, 543, 611 

\bibitem[Perlman et al.(2001)]{2001ApJ...551..206P} Perlman, E.~S., 
Biretta, J.~A., Sparks, W.~B., Macchetto, F.~D., 
\& Leahy, J.~P.\ 2001, \apj, 551, 206 

\bibitem[Perlman 
\& Wilson(2005)]{2005ApJ...627..140P} Perlman, E.~S., \& Wilson,
A.~S.\ 2005, \apj, 627, 140 

\bibitem[Perlman et al.(2007)]{2007ApJ...663..808P} Perlman, E.~S., et al.\ 
2007, \apj, 663, 808 

\bibitem[Pilbratt et al.(2010)]{Herschel} Pilbratt, G., et al.\ 2010, A\&A,
this issue

\bibitem[Poglitsch et al.(2010)]{PACS} Poglitsch, A., et al.\ 2010, A\&A,
  this issue

\bibitem[R{\"o}ser \& Meisenheimer(1999)]{1999LNP...530.....R}
  R{\"o}ser, H.-J., \& Meisenheimer, K.\ 1999, The Radio Galaxy
  Messier 87, 530,

\bibitem[Salom{\'e} \& Combes(2008)]{2008A+A...489..101S} Salom{\'e},
  P., \& Combes, F.\ 2008, \aap, 489, 101

\bibitem[Shi et al.(2007)]{2007ApJ...655..781S} Shi, Y., et al.\ 2007,
  \apj, 655, 781

\bibitem[Simionescu et al.(2008)]{2008A+A...482...97S} Simionescu, A.,
  et al.\ 2008, \aap, 482, 97

\bibitem[Sparks et al.(1993)]{1993ApJ...413..531S} Sparks, W.~B., Ford, 
H.~C., \& Kinney, A.~L.\ 1993, \apj, 413, 531 

\bibitem[Sparks et al.(2004)]{2004ApJ...607..294S} Sparks, W.~B., et
  al.\ 2004, \apj, 607, 294

\bibitem[Swinyard et al.(2010)]{Swinyard} Swinyard B., Ade P.,
  Baluteau J.-P., et al.\ 2010, A\&A, this issue

\bibitem[Tan et al.(2008)]{2008ApJ...689..775T} Tan, J.~C., Beuther, H., 
Walter, F., \& Blackman, E.~G.\ 2008, \apj, 689, 775 

\bibitem[Xilouris et al.(2004)]{2004A+A...416...41X} Xilouris, E.~M.,
  et al.\ 2004, \aap, 416, 41

\bibitem[Werner et al.(2010)]{2010arXiv1003.5334W} Werner, N., et al.\
  2010, arXiv:1003.5334

\bibitem[White et al.(1997)]{1997ApJ...475..479W} White, R.~L., Becker, 
R.~H., Helfand, D.~J., \& Gregg, M.~D.\ 1997, \apj, 475, 479 

\bibitem[Wright et al.(2009)]{2009ApJS..180..283W} Wright, E.~L., et al.\ 
2009, \apjs, 180, 283 

\bibitem[Young et al.(2009)]{2009AJ....137.3053Y} Young, L.~M., Bendo, 
G.~J., \& Lucero, D.~M.\ 2009, \aj, 137, 3053 

\end{thebibliography}
\end{document}